\documentclass[12pt,preprint]{aastex}

\usepackage{graphics,graphicx}

\def \yskip{\penalty-50\vskip3pt plus 3pt minus 2pt}
\def \reference{\par \yskip \noindent \hangindent .4in \hangafter 1}
\def \abc#1#2#3#4 {\reference#1, {\sl#2}, {\bf#3}, #4}
\def \blank {\lower 5pt\hbox to 0.75in{\hrulefill}}

\def \lesssim{\mathrel{<\kern-1.0em\lower0.9ex\hbox{$\sim$}}}
\def \gtrsim{\mathrel{>\kern-1.0em\lower0.9ex\hbox{$\sim$}}}

\begin{document}

\title{ROSAT X-ray Spectral Properties of Nearby Young
Associations: \\ TW Hydrae, Tucana-Horologium, and the
$\beta$ Pic Moving Group} 

\author{Joel H. Kastner\altaffilmark{1}, Lara
Crigger\altaffilmark{1}, 
Margaret Rich\altaffilmark{1,2},
and David A. Weintraub\altaffilmark{3}}
\altaffiltext{1}{Chester F. Carlson Center for Imaging
Science, Rochester Institute of Technology, 54 Lomb Memorial
Dr., Rochester, NY, 14623, U.S.A.; JHK's email: jhk@cis.rit.edu}
\altaffiltext{2}{Rush-Henrietta Senior High School, 1799 Lehigh Station Rd.,
Henrietta, NY, 14467, U.S.A.}
\altaffiltext{3}{Dept.\ of Physics and Astronomy, Vanderbilt University,
Nashville, TN, 37235, U.S.A.; david.weintraub@vanderbilt.edu}

\begin{abstract}
We present archival ROSAT data for three recently
identified, nearby ($D<70$ pc), young ($\sim10-40$ Myr)
stellar associations: the TW Hydrae Association, the Tucana-Horologium
Association, and the $\beta$ Pic Moving Group. The
distributions of ROSAT X-ray hardness ratios (HR1, HR2) for
these three groups, whose membership is dominated by
low-mass, weak-lined T Tauri stars, are tightly clustered
and very similar to one another. The value of HR1 for TW Hya
itself --- the only {\it bona fide} classical T Tauri star
in any of the nearby groups --- is clearly anomalous among
these nearby young stars. We compare the hardness ratio
distributions of stars in the three nearby groups with those
of T Tauri stars, the Hyades, and main sequence dwarfs in
the field. This comparison demonstrates that the X-ray spectra
of F through M stars soften with age, and that F
and G stars evolve more rapidly in X-ray spectral hardness
than do K and M stars. It is as yet unclear
whether this trend can be attributed to age-dependent
changes in the intrinsic X-ray spectra of stars of type F and later,
to a decrease in the column density of circumstellar gas
(e.g., in residual protoplanetary disks), or to the
diminishing contributions of star-disk interactions to X-ray
emission. Regardless, these results demonstrate that
analysis of archival ROSAT X-ray spectral data can help both
to identify nearby, young associations and to ascertain the
X-ray emission properties of members of known associations.
\end{abstract}

\keywords{stars: Xrays --- stars: T Tauri: individual (TW
Hya) --- stars: evolution --- open clusters and
associations: individual (TW Hya Association, Tucana
Association, Horologium
Association, $\beta$ Pic Moving Group)} 

\section{Introduction}

The recent discovery of several groups of young ($\sim5-40$
Myr) stars within 100 pc of the Sun has given new direction
to the field of star and planet formation (Jayawardhana \&
Greene 2001). The seminal nearby, young stellar group is the
TW Hydrae Association (TWA), which lies only $\sim50$ pc
from Earth (Kastner et al.\ 1997), is $\sim5-10$ Myr old
(Weintraub et al.\ 2000), and has $\sim20$ known member star
systems (Webb et al.\ 1999; Zuckerman et al.\ 2001c
[hereafter Zc]). Several dozen additional, recently
identified TWA candidate systems are listed in Makarov \&
Fabricius (2001) and Webb (2001). Other examples are the
very nearby $\beta$ Pictoris Moving Group (bPMG; $D \sim 36$
pc, age $\sim12$ Myr; Zuckerman et al.\ 2001a [hereafter
Za]; Ortega et al.\ 2002), and the Tucana and Horologium
Associations (each $D \sim 40$ pc, age $\sim30$ Myr; Torres
et al.\ 2000; Zuckerman \& Webb 2000; Zuckerman et al.\
2001b [hereafter Zb]). Each of these groups consists of
$\sim20$ known and candidate member star systems. Based on
their adjacent positions and their similar distances, space
motions, and ages, Zb and de la Reza et al.\ (2001) have
proposed that the Tucana and Horologium Associations
constitute a single, young stellar group. We adopt this
suggestion in the remainder of this paper, and we designate
the combined group as the Tucana-Horologium Association
(T-HA).
 
In many respects systems such as the TWA, T-HA, and bPMG,
though only recently identified and still in a rapid state
of flux, are better suited to detailed studies of star and
planet formation than more distant, well-studied
star-forming regions like the Orion and Taurus molecular
clouds.  Unlike T Tauri stars in regions of active star
formation, the nearby groups typically are not readily
associated with parent molecular clouds and, as a result,
there remains considerable uncertainty concerning their
origin and evolutionary status (see reviews in Jayawardhana
\& Greene 2001).  Indeed, the approximate age range of the
TWA stars, and consequently the identification of this group
as a nearby association, was initially ascertained by
Kastner et al.\ (1997) largely through the strength of the
stars' X-ray emission.  The TWA's estimated age of about 10
Myr makes the association especially intriguing, since this
age corresponds to the epoch of Jovian planet formation,
according to present theory, and is a defining
characteristic of post-T Tauri stars (Herbig 1978). The
stars in the TWA and other nearby young associations,
therefore, likely represent a long-sought missing link
between the very young and well-studied T Tauri stage and
the stellar main sequence (Jensen 2001).

As a consequence of the proximity of the TWA, X-ray observations of
its members (and of TW Hya in particular) have yielded new insight
into the origin of high-energy emission from young stars.  
Archival data from the Roentgen Satellite (ROSAT) demonstrate that the
$\sim10$ Myr age of the TWA represents an especially X-ray-luminous epoch in
the early evolution of solar-mass stars (Kastner et al.\ 1997), and
X-ray (ASCA and ROSAT) spectral monitoring suggests the
optical and X-ray variability of
some classical T Tauri stars is due to a combination
of short-term flaring and long-term variations in absorbing column
(Kastner et al.\ 1999).  Chandra/HETGS (High Energy
Transmission Gratings Spectrograph) observations of TW Hya itself
produced the first high-resolution X-ray spectrum of a T Tauri star
(Kastner et al.\ 2002).  The HETG spectra yield unexpected results for
plasma density, temperature, and elemental abundances which, taken together,
suggest that the X-ray emission from TW Hya may arise
from accretion rather than from coronal activity. If so, this would
suggest that many other X-ray luminous young stars derive part or
most of their X-ray luminosity from accretion or other
star-disk interactions, although coronal
activity remains the widely accepted mechanism for such emission (e.g.,
Feigelson \& Montmerle 1999).

Motivated by these results, we are conducting an archival study of
ROSAT data that focuses on the TWA and other nearby stellar
groups. Our goal is to establish the gross X-ray spectral properties
of nearby post-T Tauri stars. Such data, when compared with similar
results already available for T Tauri stars embedded in molecular
clouds (e.g., Neuhauser et al.\ 1995) and for main-sequence field
stars (e.g., Fleming et al.\ 1995), should offer clues to
the evolutionary status of stars 
in nearby associations and, more generally, to the mechanism(s)
responsible for the bright X-ray emission that appears ubiquitous
among young (age $\le 100$ Myr), solar-mass stars. Here, we present an
analysis of ROSAT Position-Sensitive Proportional Counter data
available for the known members of the TWA, the TA, and the bPMG. These
results strengthen the interpretation that the 
stars in these and other nearby, dispersed young associations
constitute a transition stage between cloud-embedded T Tauri stars and
main sequence stars.

\section{Data}

Data presented here consist of ROSAT
Position-Sensitive Proportional Counter (PSPC) X-ray
hardness ratios, as catalogued in the RASS Bright
Source Catalog (BSC; Voges et al.\ 1999)\footnote{These ROSAT data
were obtained through the 
the High Energy Astrophysics Science Archive Research Center, a service of
the Laboratory for High Energy Astrophysics at NASA/GSFC and 
the High Energy Astrophysics Division of the Smithsonian
Astrophysical Observatory.}. The ROSAT PSPC was sensitive
from about 0.1 to 2.4 keV. Despite the limited spectral
resolution of the PSPC ($E/\Delta E \sim 1$), 
analysis of PSPC data, and hardness
ratios in particular, has proven to be an effective means to
understand the X-ray spectral properties of young stars
(e.g., Neuhauser et al.\ 1995). As described in Neuhauser et al.,
PSPC hardness ratios (hereafter HRs) are constructed
from the integrated counts in three different PSPC
energy bands (``soft,'' ``hard 1,'' and ``hard 2,'' respectively),
spanning the energy (channel) ranges 0.1--0.4 keV (11--41), 
0.4--0.9 keV (52--90), and 0.9--2.0 keV (91--201). Denoting the
counts in these three bands as $S$, $H1$, and $H2$,
respectively, the two standard ROSAT PSPC HRs are
then defined as  
$$
HR1 = \frac{H1 + H2 - S}{H1 + H2 + S}
$$
$$
HR2 = \frac{H2 - H1}{H2 + H1}.
$$

\section{Results}

\subsection{Hardness Ratios of Local Associations}

In Fig.\ 1a, we present PSPC HRs for established members of
the TWA, as listed in Zc. With the exception of TW Hya
itself, the PSPC HRs of the TWA stars are
evidently rather tightly confined. 
It is apparent from Fig.\ 1a that, while HR2 for
TW Hya itself is typical of TWA stars, the value of
HR1 for TW Hya is significantly different from the
HR1 values of most other TWA members. This distinction in HR1 is
consistent with the unusual ultraviolet through radio
spectral properties of TW Hya. Indeed, TW Hya remains the
only unambiguous example of a classical T Tauri star among
the known TWA membership.

The membership of the T-HA remains remains subject to
debate. Here, we adopt the membership proposed by Zb and Za.
Specifically, we include all Tucana candidate members in Tables 1
and 2 of Zb except for HIP 92024, which was subsumed into
the bPMG by Za, and all but five Horologium candidate
members listed in Table 5 of Torres et al.\ (2000). The
rejected five stars (ERX 14, 16, 45, 49, and 5) were judged
by Zb to be too distant to be part of the T-HA; including
these five stars in our HR sample would not significantly
change the results described below. Seven
additional T-HA candidate members proposed by Torres et al.\
and Zb, including the infrared-excess object HD 10647, were
not detected in the RASS. Most of these stars are of
sufficiently early type (F5 and earlier) that they may be
intrinsically X-ray-faint (see below). The lack of a
RASS X-ray counterpart to the F9 star HD 10647 casts some doubt on its
membership in the T-HA, however, given the RASS detections of
several other T-HA and bPMG stars of similar spectral type.

With its membership thus defined, the distribution of HRs
for the T-HA is quite well confined, and the locus of HRs
for the T-HA lies very near that of the TWA (Fig.\ 1b). The
same is true for the bPMG (Fig.\ 1c), for which we include all
probable and possible members identified by Za. 

In Table 1 we list the
mean HRs (and errors on the means) for the TWA, the T-HA,
and the bPMG.
Aside from HR 4796A (whose M-type companion,
HR 4796B, is most likely the X-ray source in this binary
system; Jura et al.\ 1998), the known membership
of the TWA consists exclusively of K and M stars. The T-HA and
bPMG memnership includes a larger proportion of earlier-type stars, and
for these two associations we also list separately in Table
1 the mean HRs for K and M stars and for earlier types.  In calculating the 
mean HRs for the earlier-type stars, we have omitted the
handful of A stars that are associated with X-ray sources,
since it is likely that the X-rays from such systems
originate from later-type companions (e.g., Daniel et al.\
2002).

It is clear from Fig.\ 1 and Table 1 that all three nearby, young
stellar groups considered here display very similar X-ray
spectral properties in the RASS data. There appears to be a
weak tendency, moreover, for the earlier-type (F and G) stars in these young
associations to exhibit harder X-ray emission, as measured
by HR1, than the later-type (K and M) stars. 

\subsection{Comparison with T Tauri Stars, the Hyades, and Field Main
Sequence Stars}

In Fig.\ 2 we compare the HRs of the TWA, T-HA, and bPMG
members with the HRs of classical T Tauri stars (cTTS) and
weak-lined T Tauri stars 
(hereafter wTTS) in Taurus (Neuhauser et al.\ 1995), with
Hyades members detected in the ROSAT All-Sky Survey (Stern,
Schmitt, \& Kahabka 1995) for which data are available in
the BSC, and with nearby field main sequence stars. The last
sample consists of the K and M dwarfs studied by Fleming et
al.\ (1995; $D < 7$ pc, ages $\stackrel{>}{\sim}$ 1 Gyr) and
all single F and G stars  
within 14 pc\footnote{As determined through a search of the Nstars
database, http://nstars.arc.nasa.gov/index.cfm.} for
which data 
are available in the BSC. Aside from 
$\pi^1$ UMa, which likely is a member of the Ursa Major
moving group (age $300$ Myr; e.g., Messina \& Guinan
2002), the stars 
in the nearby F-G sample likely have ages of at least 2 Gyr
(e.g., Habing et al.\ 2001).
In Table 1 and Fig.\ 3 we display the mean HRs for the Taurus TTS,
local association, Hyades, and field star samples; in calculating these
means and their formal errors, we use as weights the inverse squared
uncertainties in HR1 and HR2. 

It is readily apparent from Figs.\ 2 and 3 that the typical HRs for
the TWA, T-HA, and bPMG, which cluster in the vicinity of HR1
$=0$, HR2 $=0$, differ significantly from the HRs of cTTS
and wTTS associated with the Taurus star-forming
clouds and from the field dwarf population. 
The cloud TTS have larger values of both HR1 and
HR2 than the TWA, T-HA, and pBMG stars. The wTTS in Taurus
display HR1 in the range $0.8-1.0$, while cTTS in Taurus all
display HR1 $=1.0$. The hardness ratios of field dwarfs, on
the other hand, are centered well to
the left of and somewhat below the hardness ratios of the
nearby associations; all of these stars have negative values
of HR1, and the vast majority have negative values of
HR2. Thus, although there is some overlap between the 
three populations (Fig.\ 2) --- cloud TTS, nearby young association,
and field main sequence --- Figs.\ 2 and 3 indicate that they
represent a sequence of decreasing HR. 

The distinction between the HR distributions of the members of
nearby young associations and the Hyades is more subtle
(Figs.\ 2, 3). There is substantial
overlap between the HR distributions of these two samples
(Fig.\ 2). The 93 Hyades members in the RASS are all confined to HR1 $<
0.2$, however, whereas 7 of 53 (13\%) of known members of nearby young
associations display HR1 $> 0.2$. Consistent with this
observation, the means of both HR1 and HR2 for the Hyades
stars are significantly smaller than the means of the
members of nearby young associations (Fig.\ 3, left panel). It is
apparent, however, that these distinctions are primarily the
result of the sharp distinction between the mean HRs of the
F and G stars in the young association and Hyades samples
(Fig.\ 3, right panel). For K and M stars, there is no
statistical difference between the HRs of members of nearby,
young associations and those of the Hyades members (Fig.\ 3,
center panel).

There is also a significant difference between the mean HR1
values of the Hyades and main sequence field populations,
for both F,G and K,M stars. However, the weighted mean of HR1 for the
field star F,G sample is strongly influenced by the nearest
G star, $\alpha$ Cen, because the uncertainties in its HR values
($\pm0.01$) are much smaller than those of the remainder of
the field F and G star sample (which range from $\sim\pm0.05$
to $\sim\pm0.2$). Excluding this star, whose values of HR1
$=-0.98$, HR2 $=-0.38$ place it at the
extreme left of the field star HR distribution, the the F,G star HR
distributions of the Hyades and the field stars appear
quite similar (Fig.\ 2).

\section{Discussion}

\subsection{Does X-ray emission soften with stellar age?}

Figs.\ 2 and 3 suggest
that X-ray emission softens monotonically (toward smaller values of
HR1 and, to a lesser extent, HR2) as stars evolve
from cloud T associations to sparse,
cloud-less T associations (as represented by the TWA, T-HA,
and bPMG stars), then again from cloud-less T association
to aging cluster (Hyades) and, finally, from cluster to the field. 
Such a trend of softening
X-ray radiation with stellar age has been noted by
previous investigators, based on observations of TTS and
open clusters (including the Hyades; e.g., 
Neuhauser 1997). This trend is much more clearly
demonstrated herein --- through the inclusion of the newly
identified local associations --- than in the previous work. 
We now consider the possible origin and interpretations of
such a trend.

\subsubsection{Distance, absorption, and hardness ratio}

The Hyades ($D \sim 45$ pc; e.g., Gunn et al.\ 1988) and the
local associations lie at very similar distances, such that
any systematic differences in ROSAT HRs most likely are not
due to differences in intervening interstellar
absorption. We must
consider, however, whether the trend evident in Figs.\ 2 and 3 partly  
represents monotonically decreasing interstellar absorption,
from the most distant sample (cloud 
TTS) to the least distant sample (nearby, field main
sequence stars). One does expect
objects of a given X-ray emission temperature ($T_x$) to
move almost directly to the right (increasing HR1) in the
hardness ratio diagram as absorbing column ($N_H$)
increases (see Fig.\ 4 of Neuhauser et al.\ 1995). However,
for $T_x \sim 25$ MK, HR1 only
increases from $\sim0.2$ to $\sim0.5$ as $N_H$ increases by
2 orders of magnitude (from $10^{18}$ cm$^{-2}$ to $10^{20}$
cm$^{-2}$), while HR2 remains roughly constant (Neuhauser et al.). 
Thus, while the typical extinction (and hence $N_H$) toward
stars in local associations remains to be
determined\footnote{Rucinsky \& Krautter (1983) noted that 
foreground extinction is negligible in the general
direction of TW Hya itself.}, it
is likely that the local ISM gas column densities are
not large enough to explain the differences in
mean HR1 between local, older (age $\stackrel{>}{\sim} 2$
Gyr) field stars and the younger stars in the Hyades 
and in local groups such as the TWA, T-HA, and bPMG. 
Furthermore, we find no statistical difference
between the HRs of TWA members recently identified
by Zc and those of the ``original'' TWA members identified
by Kastner et al.\ (1997) and Webb et al.\ (1999), despite
the fact that the former stars are generally more distant than the
latter (Zc). The trend apparent in Figure 2 therefore is
caused predominantly by age and/or circumstellar environment (\S 5).

The situation may be somewhat more complicated for samples
that lie as distant as Taurus ($D \approx 140$ pc), as at such distances
interstellar gas may have a significant affect on HR1.
Indeed, the effects of interstellar absorption may explain
the positions of Pleiades stars in a plot of 
HR1 vs.\ HR2; these $\sim100$ Myr-old stars are clustered
near HR1 $\sim0.5$ (see Fig.\ 6 of Neuhauser 1997), i.e.,
their ROSAT X-ray spectra are softer than those of cloud TTS
but are somewhat harder, on average,
than those of the TWA, T-HA, and bPMG members, particularly in
HR1. The typical reddening toward
Pleiades members, $E(B-V) \approx 0.04$ (Breger
1986), implies typical column densities $N_H \approx
2\times10^{20}$ cm$^{-2}$ (Neuhauser et al.\ 1995), so this
difference in mean HR1 can be explained as partly due to
extinction, suggesting that the intrinsic HRs of the Pleiades stars
are similar to those of the
younger TWA and other local groups.  

\subsubsection{Spectral type and hardness ratio}

Fig.\ 3 and Table 1 indicate that earlier-type (F and G)
stars evolve more rapidly in HR than do late-type (K and M)
stars. Specifically, it would appear that at the age of the
local associations ($\sim10-30$ Myr), F and G stars display
somewhat harder X-ray emission than do K and M stars, while
by Hyades age (700 Myr) the situation is reversed.  A recent
{\it Chandra} study of the Pleiades (Daniel et al.\
2002) suggests that F and G
types already display softer X-ray spectra than K and M
types by the time such stars are $\sim$100 Myr old. The
relatively rapid X-ray evolution of F and G stars (relative
to K and M types) is consistent with the result that K and M
stars reach the X-ray ``saturation'' level ($\log{L_x/L{\rm
bol}} \sim -3$) at $\sim 100$ Myr (Kastner et al.\ 1997), 
much later than F and G stars (e.g., Randich et
al.\ 1995). While Fig.\ 3 indicates that K and M
stars evolve very little in X-ray spectral hardness between
$\sim10$ Myr and 700 Myr, K and M stars do become softer
X-ray sources after several Gyr.

\subsection{ROSAT hardness ratios of candidate TWA members}

Based on an optical spectral survey of TWA candidate stars
identified by Makarov \& Fabricius (2001; hereafter MF),
Song et al.\ (2002; hereafter SBZ) conclude that 20 of these
23 candidates are not TWA members. Many or most of the MF
stars appear to be $\sim100$ Myr old, based on their
positions in theoretical pre-main sequence isochrones (Fig.\
2 of SBZ) and, to a lesser extent, on the strengths of their
Li $\lambda6708$ absorption lines (Fig.\ 1 of SBZ).  Fig.\ 4
demonstrates that the HRs of most of the MF stars follow the
HR distribution of the ``established'' TWA as defined by
Zc. If the results of SBZ are correct, therefore, Fig.\ 4
further indicates that there is little evolution in the X-ray
spectral properties of late-type stars between
$\sim10$ Myr and $\sim100$ Myr. 

The only three Makarov \& Fabricius (2001) TWA candidates
whose spectral properties are consistent with TWA
membership, according 
to SBZ, display somewhat anomalous (TW Hya-like) HRs in
Fig.\ 4. The absence of strong H$\alpha$ emission in the SBZ
spectra of these three stars (TYC 7760--0835--1, TYC 8238--1462--1, and TYC
8234--2856--1) indicates that (unlike TW Hya) these stars are
not classical TTS. If these stars are indeed roughly
coeval with and have HRs similar to those of the TWA, then
their rightward positions, relative to previously
established TWA members, in Fig.\ 4 may be due in part to their 
relatively early spectral types; all three stars have $B-V$
colors that suggest they are earlier than spectral type K. 
We may also view these stars through a
larger column density of interstellar absorbing material
than most other TWA members (\S
4.1). With regard to the latter possibility, we note that SBZ 
derive photometric distances of $\sim130$ pc to TYC
7760--0835--1, TYC 8238--1462--1, and TYC 8234--2856--1 ---
i.e., these stars are roughly twice as distant as 
the ``established'' TWA membership (Zc). 

\section{Conclusions}

Our analysis of the ROSAT PSPC hardness ratios of local associations, and
of cloud T Tauri stars and field main sequence stars,
indicates that X-ray spectral hardness 
decreases monotonically with
increasing stellar age (Figs.\ 2 and 3 and Table 1), with the trend
stronger for F and G stars than for K and M stars.
There are three alternative
interpretations of this trend: 
\begin{enumerate}
\item The intrinsic X-ray spectrum of a star of spectral
type F or later softens with
age (Fleming et al.\ 1995; Neuhauser 1997). Such a trend
could be due to decreasing X-ray emission
temperature and/or to age-dependent changes in X-ray
emitting region abundances that modulate the
strengths of the brightest emission lines in the ROSAT
band. 
\item The column density of X-ray absorbing  
material associated with the environments of young stars
declines monotonically as these objects evolve. For Taurus
TTS, the X-ray absorbing gas and 
dust may consist of both circumstellar and molecular cloud
material, whereas for the older, nearby stellar groups like
the TWA, T-HA, and bPMG, we
expect residual circumstellar material to dominate. 
This would suggest that PSPC hardness ratios of the members
of local associations can
be used to examine the evolution of gaseous 
circumstellar disks around young stars. 
\item Contributions to X-ray emission from star-disk
interactions in general, and accretion in particular,
decline as stars evolve from the T Tauri phase
toward the main sequence. Such contributions may arise in
star-disk magnetic field reconnection events (e.g., Shu et
al.\ 1997) or in energetic shocks along accretion columns
(Kastner et al.\ 2002). In either model, the emitting region
temperature would be in excess of $\sim10^6$ K, and
therefore should contribute a relatively hard excess X-ray
emission component that diminishes with age as the disk disperses.
\end{enumerate}

For F and G stars --- which continue to evolve in HR from
the T Tauri through post-T Tauri through early main sequence
stages --- we cannot distinguish, at present, between these
alternative explanations. Given that TW Hya is surrounded by
a disk viewed nearly pole-on (e.g., Kastner et al.\ 2002 and
references therein), the anomalous position
of this K7 star in Fig.\ 1 offers support for the third
interpretation (declining star-disk interactions) in the case of 
young ($\stackrel{<}{\sim} 10$ Myr), low-mass stars. 
The first interpretation (changing intrinsic stellar X-ray emission
properties) best explains the
continuing X-ray spectral evolution of K and M stars beyond 
the age of the Hyades, however. We are
analyzing archival PSPC spectra of  
individual stars to determine whether further progress can
be made from the available ROSAT data (Kastner \& Crigger,
in preparation). However, high-resolution Chandra and XMM
X-ray spectroscopy of representative objects --- from which
precise X-ray emitting region temperatures, densities, and
elemental abundances can be determined (e.g., Kastner et
al.) --- will no doubt be required to establish the
mechanism(s) responsible for the softening of X-ray emission
with stellar age.

Regardless of the correct interpretation of the observed
trend, Figs.\ 2 and 3 demonstrate that analysis of ROSAT HRs can
help discriminate between X-ray-emitting T Tauri stars,
post-T Tauri stars, and both young (Hyades- and Pleiades-age)
and older (age $\stackrel{>}{\sim} 1 $ Gyr) main sequence stars in the 
field. ROSAT HRs evidently are a particularly useful tool for
assessing the ages of samples that consist mostly of
F and G stars, while for
spectral types of K and later, ROSAT HRs can effectively
isolate peculiar cases (e.g., TW Hya). The
results presented here suggest, therefore, that 
candidate members of young associations can be identified
partly on the basis of ROSAT PSPC hardness ratios,
particularly if other age-related X-ray emission
discriminants are applied (such as the ratio of X-ray
luminosity to bolometric luminosity; Kastner et al.\ 1997).

\acknowledgements{We acknowledge incisive comments from the
referee and Inseok Song that substantially improved this paper.
Ms.\ Rich acknowledges support from a summer internship program funded
by the Industrial Associates of the Center for Imaging Science at RIT.}

\begin{deluxetable}{ccrr}
\tablecaption{\sc Mean ROSAT PSPC Hardness Ratios}
\tabletypesize{\small} \tablewidth{0pt} 
\tablehead{
\colhead{Sample\tablenotemark{a}} & 
\colhead{No.\tablenotemark{b}} & 
\colhead{HR1} & \colhead{HR2} } 
\startdata 
\multicolumn{4}{c}{\sc All stars} \\
Taurus T Tauri     & 64 & $0.833\pm0.033$ & $0.289\pm0.043$ \\
TW Hya Association & 16 & $0.040\pm0.070$ & $0.025\pm0.047$ \\
$\beta$ Pic Moving Group & 15 & $-0.060\pm0.025$ & $0.054\pm0.016$ \\
Tuc-Hor Association & 32 & $-0.008\pm0.025$ & $0.015\pm0.034$ \\ 
Hyades             & 93 & $-0.234\pm0.027$ & $-0.048\pm0.028$ \\
Nearby field stars & 74 & $-0.710\pm0.045$ & $-0.129\pm0.018$ \\
\multicolumn{4}{c}{\sc K,M stars} \\
TW Hya Association & 16 & $0.040\pm0.070$ & $0.025\pm0.047$ \\
$\beta$ Pic Moving Group & 9 & $-0.081\pm0.021$ & $0.062\pm0.015$ \\
Tuc-Hor Association & 15 & $-0.057\pm0.024$ & $-0.054\pm0.043$ \\ 
Hyades             & 31 & $-0.104\pm0.027$ & $-0.011\pm0.033$ \\
Nearby field stars & 61 & $-0.544\pm0.042$ & $-0.083\pm0.021$ \\
\multicolumn{4}{c}{\sc F,G stars} \\
$\beta$ Pic Moving Group & 4 & $0.104\pm0.074$ & $0.009\pm0.045$ \\
Tuc-Hor Association & 16 & $0.029\pm0.040$ & $0.059\pm0.047$ \\ 
Hyades             & 62 & $-0.313\pm0.035$ & $-0.044\pm0.039$ \\
Nearby field stars & 13 & $-0.928\pm0.050$ & $-0.375\pm0.011$ \\
\enddata
\tablenotetext{a}{ See text for definitions and references. }
\tablenotetext{b}{ Number of X-ray emitting stars included
in calculations of means and errors on the means. }
\end{deluxetable}

\newpage


\begin{figure*}[htb]
\includegraphics[scale=0.9,angle=0]{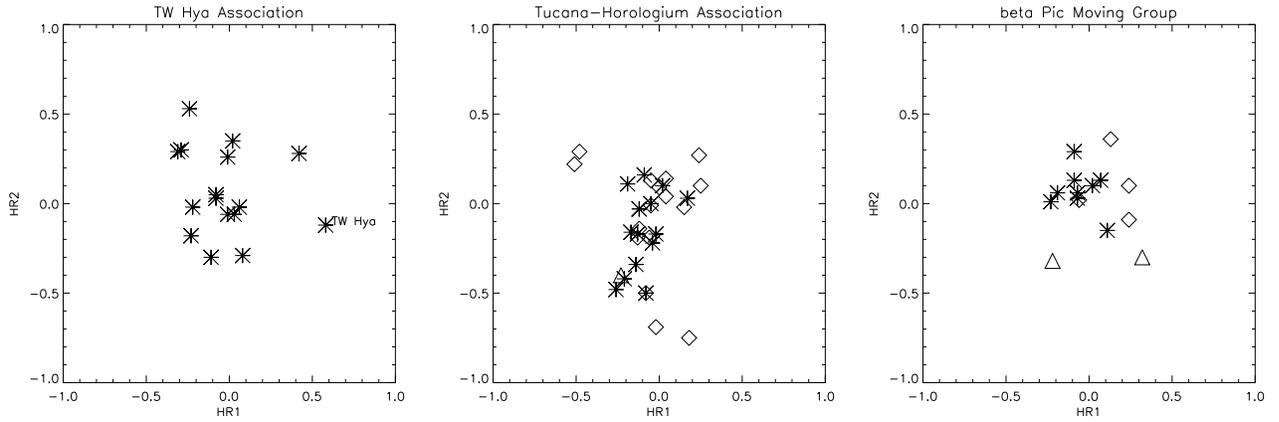}
\caption{ROSAT PSPC hardness ratios for presently established 
members of the TW Hya Association (left), the Tucana-Horologium
Association (center), and the $\beta$ Pic Moving Group
(right). The position of TW Hya is indicated in the left panel. In
each panel, K and M stars are indicated as asterisks, F and
G stars as diamonds, and A stars as triangles. Typical
uncertainties in HRs range 
from $\pm0.05$ to $\pm0.2$.}
\end{figure*}

\begin{figure*}[htb]
\includegraphics[scale=0.9,angle=0]{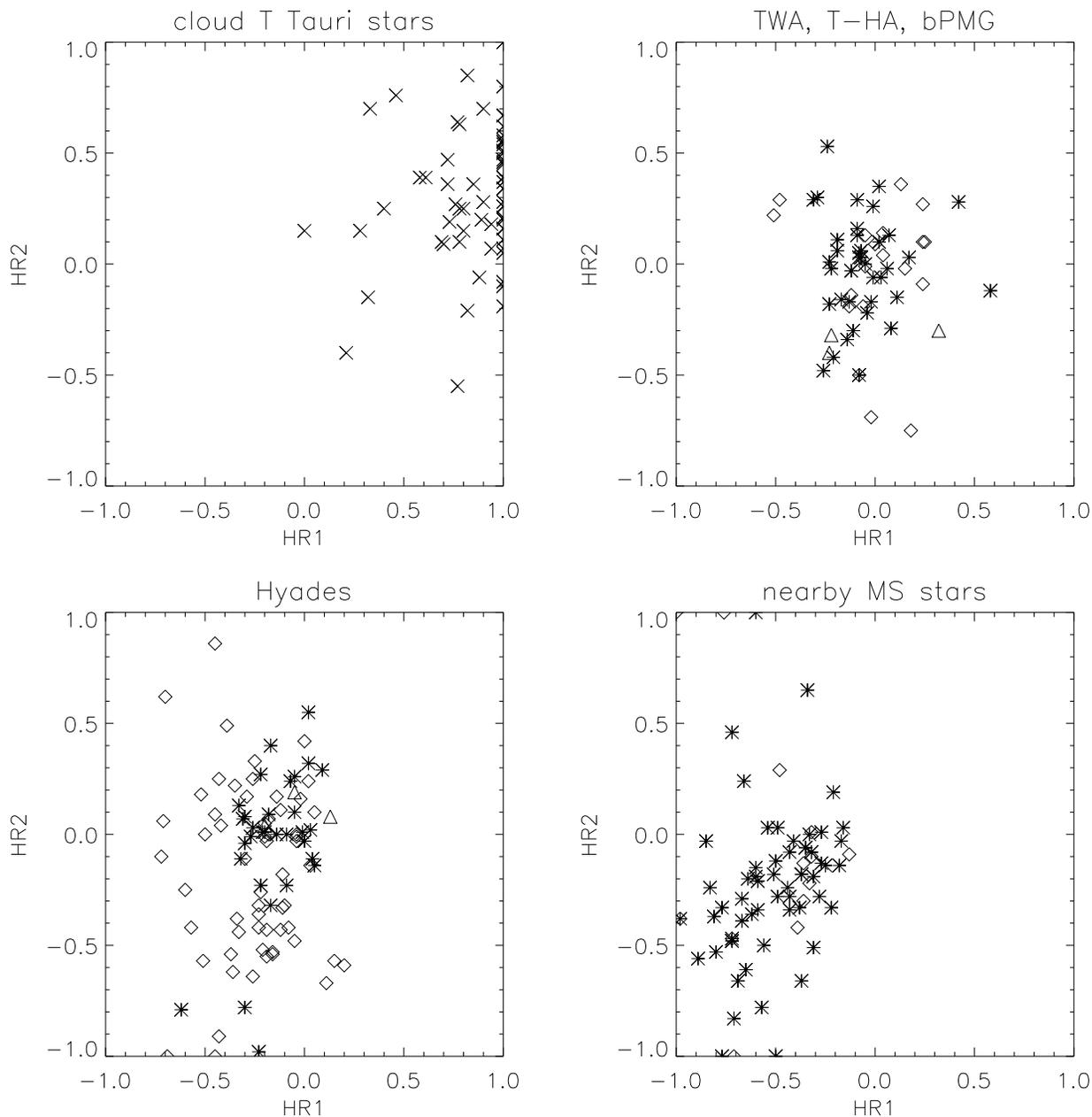}
\caption{Top left: HRs for T
Tauri stars in Taurus (Neuhauser et al.\ 1995). The points
at the extreme right of the plot are predominantly cTTS, all
of which display $HR1 = 1.0$. Top right:
HRs for established members of all three local 
associations (TWA, T-HA, $\beta$ Pic). Bottom left: HRs for
Hyades members (Stern et al.\ 1995). Bottom right: HRs
for nearby ($D < 7$ pc) field K and M dwarfs (Fleming et
al.\ 1995) and for single F and G stars within 14
pc. Symbols are as in Fig.\ 1. Typical uncertainties in HRs range
from $\pm0.05$ to $\pm0.2$.
}
\end{figure*}

\begin{figure*}[htb]
\includegraphics[scale=0.9,angle=0]{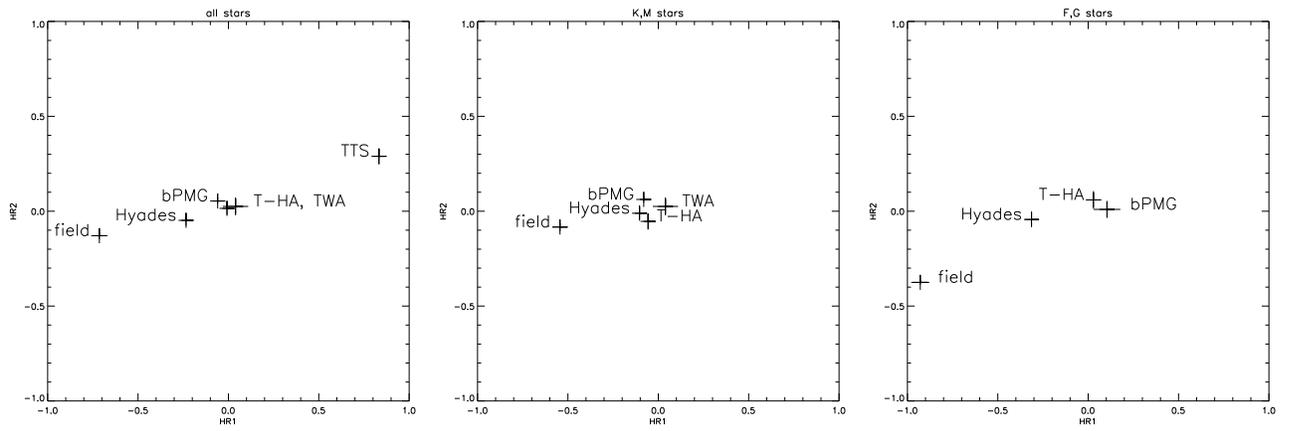}
\caption{Means and errors on the mean
for the hardness ratios of Taurus TTS (``TTS''), members of the TWA,
the T-HA, and the bPMG, and field stars. The left panel
illustrates the mean HRs for all stars, 
the center panel for K and M stars, and the right panel for
F and G stars.
}
\end{figure*}

\begin{figure*}[htb]
\includegraphics[scale=1.,angle=0]{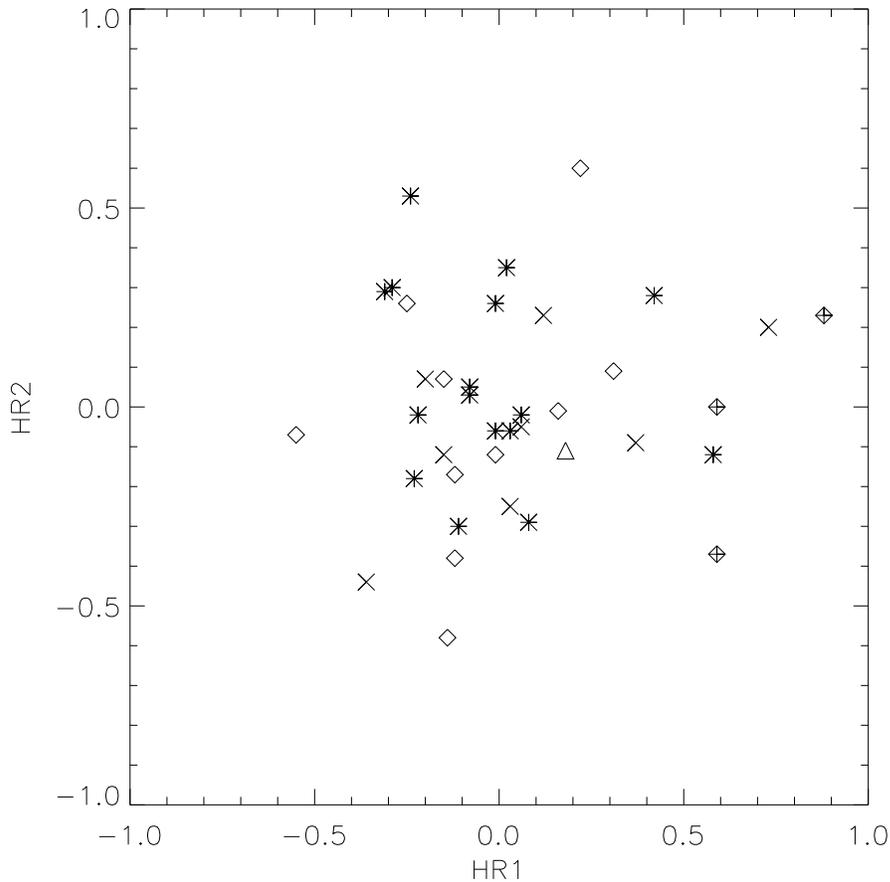}
\caption{HRs for known TWA
members (asterisks) and for candidate TWA members
identified by Makarov \& Fabricius (2001). For the Makarov
\& Fabricius stars, crosses indicate K and M stars, diamonds
F and G stars, and the triangle is an A star (where we infer
these spectral types from $B-V$ colors as listed in Song et
al.\ 2002). The only three Makarov \& Fabricius candidate
stars that Song et al.\ (2002)  
tentatively confirmed as TWA members are the three rightmost
F and G stars in the diagram (indicated with plus signs).
}
\end{figure*}

\end{document}